# A symbolic machine learning approach for cybersickness potential-cause estimation

Thiago Porcino[1], Erick O. Rodrigues[2], Flavia Bernardini[1], Daniela Trevisan[1], and Esteban Clua[1]

[1] Universidade Federal Fluminense, Niteroi, Brazil
[2] Universidade Tecnológica Federal do Paraná, Pato Branco, Brazil
thiagomp@ic.uff.br

**Abstract.** Virtual reality (VR) and head-mounted displays are constantly gaining popularity in various fields such as education, military, entertainment, and bio/medical informatics. Although such technologies provide a high sense of immersion, they can also trigger symptoms of discomfort. This condition is called cybersickness (CS) and is quite popular in recent publications in the virtual reality context. This work proposes a novel experimental analysis using symbolic machine learning that ranks potential causes for CS. We estimate the CS causes and rank them according to their impact on the classification capabilities of CS. The experiments are performed using two distinct virtual reality games. We were able to identify that acceleration triggered cybersickness more frequently in a race game in contrast to a flight game. Furthermore, participants less experienced with VR are more prone to feel discomfort and this variable has a greater impact in the race game in contrast to the flight game, where the acceleration is not controlled by the user.

## 1 Introduction

Activities such as virtual training environments, simulations and entertainment in immersive virtual formats are constantly becoming more popular with the continued development and public interest in VR technologies over the last years [7]. In 2019, the VR hardware market was valued at 4.4 billion US dollars and is expected to reach 10 billion US dollars by 2022 [32].

Head-mounted displays (HMDs) are one of the means of achieving immersive virtual reality environments. These devices usually consist of electronic displays and lenses that are fixed over the head where the display and lenses face the eyes of the user. HMDs are used for various purposes in the industry such as in games that focus on applications for numerous contexts [21].

Unfortunately, HMDs are strongly related to frequent manifestations of discomfort [20]. Among the possible manifestations, cybersickness (CS) deserves special attention as it is the most frequent and is usually associated to long exposures to HMDs.

The most frequent symptoms caused by CS are general discomfort, headache, stomach awareness, nausea, vomiting, sweating, fatigue, drowsiness, disorientation, and apathy [9]. These symptoms impact the user experience and affects the VR industry negatively.

Several works in the literature address the CS phenomenon and mitigation strategies for immersive VR applications using HMDs [28, 27]. This work estimates and is amenable to rank the attributes that contribute the most in terms of triggering the cybersickness. We propose an approach that enables this estimation in real time, while the user is under the VR condition. Decision tree algorithms are symbolic and adequate for this occasion, where understandability is essential [23].

The approach consists of following the prediction course through the decision tree while using feature frequency and the node height to estimate the influence of the attributes. The proposed approach can be used to assess specific CS causes in real-time scenarios, allowing for future approaches to draw conclusions over how to mitigate these causes based on specific user-game contexts.

This paper is organized as follows: Section 2 describes the literature review, the necessary background knowledge, which includes the types of sickness associated to VR, symbolic machine learning concepts, and related work. Section 3 describes the details of the dataset acquisition. Section 4 describes our approach that estimates potential causes of cybersickness. Section 5 contains a discussion of our results. At last, Section 6 describes the conclusion, which includes limitations of the current research and future work.

## 2 Literature Review

Motion sickness (MS) manifests due to the information divergence captured by the human sensory system in the presence of conflicts between the sensory organs that define orientation and spatial positioning. MS is defined as the discomfort felt during movement that is not related to body movement, e.g., in the case of airplane, boat, or land trips [2]. This type of discomfort also occurs in virtual environments and is called visually induced motion sickness (VIMS). Moreover, in VR, VIMS is used to show that sickness is likely to originate from the visual perception of motion originated from 3D images [3]. In contrast, VIMS that occurs during flight or drive simulators is often called simulator sickness. Overall, motion sickness can be split into two subcategories [17]: transportation sickness, which is tied to the real world and simulator sickness, which is associated to the virtual world and includes cybersickness.

CS symptoms are similar to MS symptoms: nausea, vertigo, dizziness, and upset stomach [29]. Manifestations related to VIMS or CS may originate from various causes. Some of these causes are already properly described in the literature, whereas others still need to be explored.

In a previous work [27], we created a detailed review of strategies with the main objective of minimizing CS causes. Furthermore, we have undergone a feature selection study that selects the most relevant attributes (causes) related to CS [26].

This work, in contrast, focuses on studying the most relevant attributes or features that lead to discomfort, in real-time. We also focus on investigating some approaches aiming to minimize CS in virtual environments using machine learning.

### 2.1 Symbolic Machine Learning Concepts

Given a training *dataset* $X = \{x_1, ..., x_n\}$ containing instances such that $X \in R^p, p$ being the number of attributes of the instances with corresponding labels $c(x) \in \{1, ..., L\}$, the classification problem consists of assigning a label $l \in \{1, ..., L\}$ to unlabelled instances $y$, assuming that $X$ can assist the process [30].

This work is built upon the understandability of *symbolic classifiers*, whose decision process description can be represented by a set of unordered or disjoint rules. Symbolic classifiers are not novel and they have been used in many scenarios where clear logical comprehensiveness is required [1]. A decision tree can be written as a set of disjoint unordered rules [11, 10].

When a decision tree is built, the set of instances in a decision tree leaf node is the one covered by the rule formed by the path starting in the root node up to the leaf node. A decision tree is illustrated in the left side of Figure 1. Moreover, a random forest, illustrated on the right side of Figure 1, is a classifier represented by a collection of decision trees. Each tree is constructed using a sub-sample of features (bagging), randomly selected from the feature domain. Each tree of the forest votes for a class during the classification phase. Usually, the mode represents the chosen class and the final decision [5].

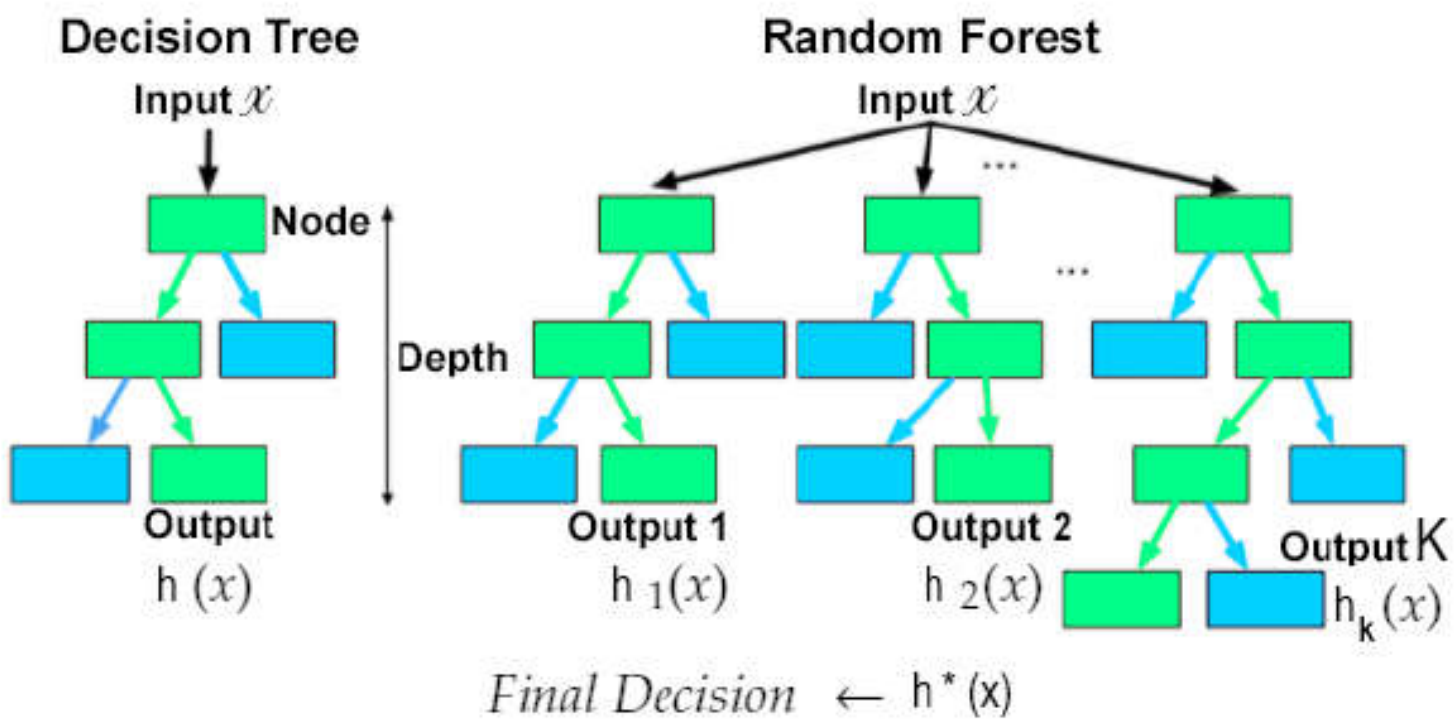

**Fig. 1.** Decision tree and random forest. The green paths represent the outcome for this particular case.

### 2.2 Related Work

Kim et al. [19] proposed a deep learning architecture to estimate the cognitive state using brain signals and how they relate to CS levels. Their approach is based on deep learning approaches such as long short-term memory (LSTM) and CNN [22, 13, 31]. The models learn the individual characteristics of the participants that lead to the manifestation of CS symptoms when watching a VR video.

Jin et al. [16] grouped CS causes as: hardware-related (e.g., tied to VR device settings and features), software-related (e.g., tied to the content of the VR scenes), and user-related. The authors used classifiers to estimate the level of discomfort. The LSTM-RNN obtained the best results.

Jeong et al. [15] focus on 360 VR streaming. The authors analyzed the scenarios where CS is associated to brain signals. Their work uses data from 24 participants to recognize the common characteristics among their VR streams and related CS manifestation. They examined the VR content and separated the segments where several individuals felt discomfort. Still, authors were not able to find specific and related individual CS causes.

Furthermore, Kim et al. [19] and Jeong et al. [15] capture data using external medical equipment. However, medical equipment is not mainstream in terms of VR. For this reason, this work proposed a framework based on data that can be captured with no specific accessories.

Garcia-Agundez et al. [12] focus on the classification of the level of CS. The proposed model uses a combination of bio-signal and game settings. User signals such as respiratory and skin conductivity of 66 participants were collected. Authors obtained a classification accuracy of 82% (SVM) when it comes to binary classifications and 56% (kNN) for the ternary case.

Moreover, Porcino et al. [26] proposed an approach that is amenable to the prediction of CS during the gameplay. Authors were able to achieve an average accuracy of 96.54% with random forest considering a total of 16 different machine learning models and different scenarios. Additionally, they identified attributes responsible for painful states in VR games.

However, Garcia-Agundez et al. [12] and Porcino et al. [26] do not attempt to estimate the weight or the influence of the attributes (i.e., the cause) leading to CS, as opposed to the proposal of this work.

In this novel approach, we are able to use symbolic machine learning to analyse and identify one or more causes of discomfort, which is user and context specific. In other words, the approach described in this manuscript is not a general rule for recognizing the presence of discomfort as previously approached in the current literature. In contrast, it provides real-time user and context-sensitive evaluation and estimation of causes for cybersickness.

Moreover, the use of symbolic classifiers is paramount for an appropriate analysis and understanding of the decision, as opposed to deep learning approaches, which are black boxes.

## 3 Materials and Methods

We used Unity 3D [8] to create two different VR games: (1) a race game and (2) a flight game. In our pipeline, we require the participants to fill in questionnaires (CSPQ [26] and two VRSQ [18]) before and after the gameplay and participate for 5 minutes in basic VR game (using Oculus Rift and HTC Vive) that contains visual movements such as rotation and translation.

In the race game occasion, the acceleration varies according to the choice of the user (they push the acceleration according to their will). In contrast, the flight game simulates an almost-constant acceleration. The player experience for both games is highlighted in Figure 2.

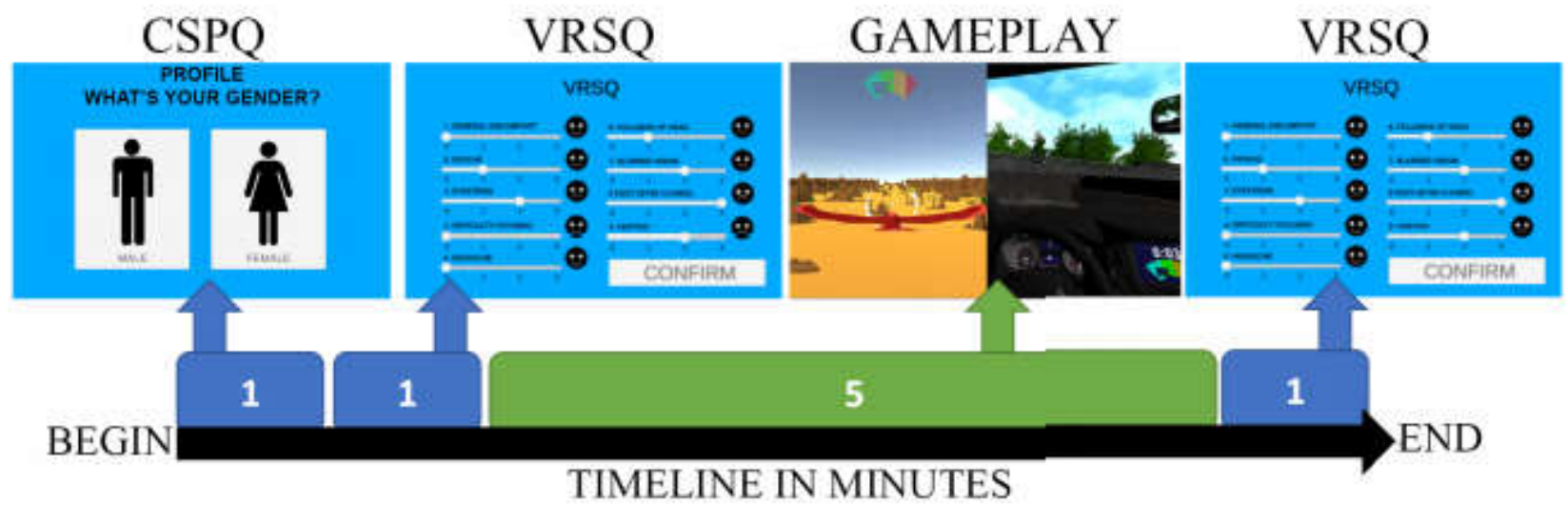

**Fig. 2.** Steps of our methodology.

A total of 35 users (9 women, 26 men) with ages ranging between 18 and 60 answered all the questionnaires correctly and completed the whole game interaction. All participants agreed with their anonymous participation in the study and signed consent forms. The participants were allowed to quit the experiment whenever they wanted. Each individual was required to complete four tasks, as follows:

- fill in the profile questionnaire (CSPQ)[26]. This questionnaire considers gender, age, previous experience with virtual environments, flicker sensitivity, any pre-symptom (such as stomach pain, flu, stress, hangover, headache, visual fatigue, lack of sleep or respiratory diseases), any vision impairment, presence of eyeglasses, posture (seated or standing) and eye dominance.
- fill in the VRSQ questionnaire [18];
- participate in one of the VR games for up to 5 min while mentioning the numbers 0 (none), 1 (slight), 2 (moderate), or 3 (severe) for each time their level of discomfort changed during the gameplay experience;
- and, at last, fill in the VRSQ questionnaire after the experience.

In terms of VRSQ results, 7/15 users from the race game scored positive for CS. When it comes to the flight game, 8/20 users reported discomfort. These scores represent 46.7% and 40.0%, respectively. All the features were captured considering two aspects: type of hardware and type of game.

The complete list of recorded parameters can be found in a previous publication [26]. We merged the top six features for each one of the scenarios (race and flight games). In addition, we discarded the time stamp and position attributes. The position attribute is specific to each game and can produce overfitting or even end up producing low accuracy rates when used with games other than the ones used during the training phase. In a similar sense, the timestamp feature is indirectly associated, in different extents, to nearly all the other features. Besides being redundant, this can also produce overfitting to the context of a specific game rather than working with generalizations. Therefore, extracted features (Table 1) were used as training data to construct the decision trees.

**Table 1.** Extracted feature set

| Gameplay data | Profile data |
|---|---|
| Speed | Gender |
| Acceleration | Age |
| Rotation Z | VR Experience |
| Frame Rate | Discomfort Label |

In summary, the four original classes (Discomfort Label in Table 1), namely, none, slight, moderate, and severe, were converted to binary: 0 for no discomfort and 1 for discomfort. Class 1 translates to slight, moderate, or severe classes. Throughout the work, we use the random forest and decision tree algorithms from the scikit-learn [24] Python library (Python version: 3.7.5 , sci-kit learn version: 0.22.1). In the next section, we provide our cybersickness cause estimation methodology.

## 4 Potential-cause estimation approach

The logical prediction path of the decision trees inherits a personal fingerprint associated to attribute weights. Usually, attributes that are closer to the tree root are more important, as they often reduce the chaos in the data more than other attributes (information gain, less entropy). As a general rule, the frequency in which attributes appear in the decision path is also an important piece of information. We combine these two aspects to estimate the importance of the attribute (i.e., the most important causes of discomfort).

Let us suppose a decision tree described by 9 decision nodes, as shown in Figure 3, which contains in their conditions the features G (gender), R (rotation Z), A (acceleration) and S (speed). Furthermore, let us consider a path for an instance that was predicted as discomfort, highlighted in green in Figure 3.

We compute a potential-cause score (PCS) by summing up the heights of these features in the green path. In this case, G appeared once and has height 3, rotation R has height 2 and speed S has heights 1 and 0. Next, the output is divided by the sum of all depths of the tree, as follows: G=3/6 or 0.5, R=2/6 or approximately 0.33 and S=1/6 or approximately 0.17. In this case, we estimate gender as the most relevant cause for CS (Equation 1)

$$PCS(F) = \frac{\sum_{h=0}^{max\ height} \{ h\ ,\ if\ F\ belongs\ to\ the\ height.\ 0,\ otherwise.}{\sum_{h=0}^{max\ height} h} \tag{1}$$

where $h$ varies from 0 to the maximal tree height, and $F$ is the feature being evaluated. PCS is computed considering just the decision path (e.g., the one highlighted in green).

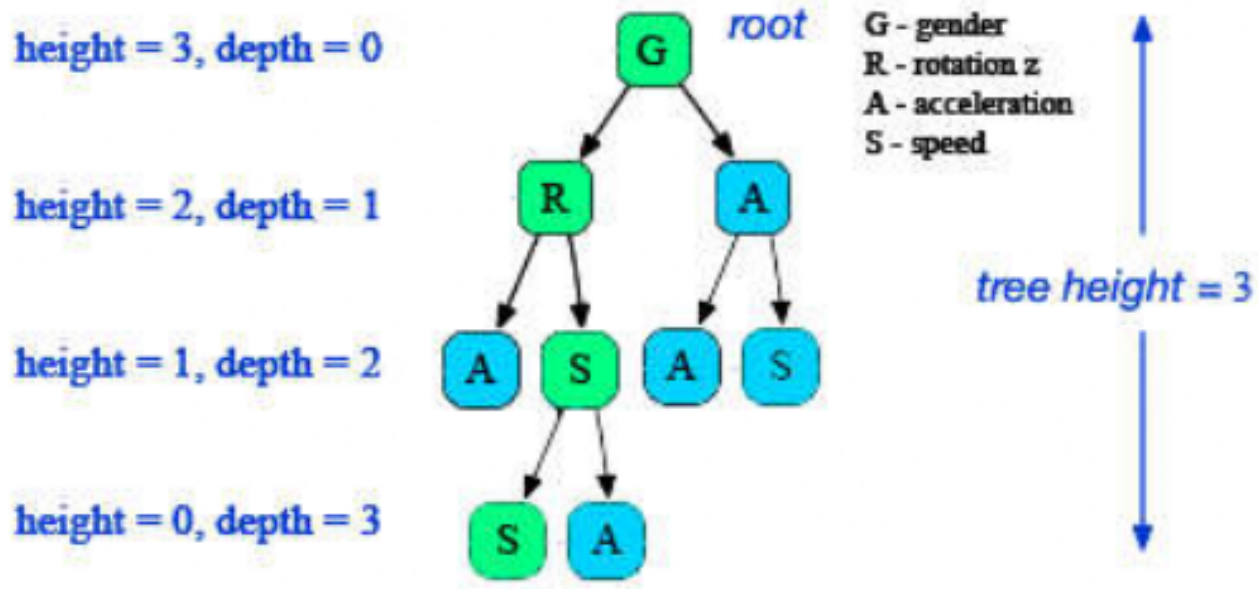

**Fig. 3.** A decision tree model. The green path illustrates the decision path that uses the attributes G, R, S, and S. Each attribute is associated to height values 3, 2, 1, and 0, respectively.

Furthermore, the random forest model can be considered a set of decision trees. We sum the PCS results from each tree $t$ if the tree final decision is equal to the RF final decision. Otherwise, we sum 0 in this iteration.

Besides, as the evaluation method, we chose to use the leave-one-user-out method, a particular case of cross-validation where the number of folds matches the number of instances in the dataset. Consequently, the algorithm is trained $n$ times, where $n$ is the total amount of participants. Additionally, we compare AUC (area under the curve) [14] scores obtained results from the random forest and the simple decision tree in terms of predictability.

When it comes to the comparative results concerning the AUC scores the random forest classifier obtained the similar results to the decision tree (Table 2 and Table 3). However, random forest is an ensemble algorithm, and hence it is naturally more complex and slower in terms of run time when compared to a single decision tree. Simple decision trees are faster and more compatible with real-time predictions, and hence we adopted them to generate the PCS feature ranking (Figure 4) for the race and flight.

**Table 2.** User-specific AUC scores in the race game (tree depth 7).

| User | Decision Tree | | Random Forest | |
|---|---|---|---|---|
| | Train | Test | Train | Test |
| 21 | 0.86 | 0.49 | 0.87 | 0.58 |
| **22** | **0.86** | **0.98** | **0.88** | **1.00** |
| 23 | 0.86 | 0.62 | 0.88 | 0.65 |
| 24 | 0.88 | 0.51 | 0.89 | 0.72 |
| 25 | 0.88 | 0.56 | 0.89 | 0.57 |
| 26 | 0.86 | 0.53 | 0.87 | 0.56 |
| 27 | 0.84 | 0.83 | 0.86 | 0.87 |
| 28 | 0.86 | 0.48 | 0.88 | 0.39 |
| 29 | 0.86 | 0.39 | 0.88 | 0.69 |
| 30 | 0.90 | 0.66 | 0.90 | 0.79 |
| 31 | 0.88 | 0.58 | 0.89 | 0.54 |
| 32 | 0.84 | 0.11 | 0.86 | 0.12 |
| 33 | 0.87 | 0.50 | 0.89 | 0.58 |
| 34 | 0.88 | 0.48 | 0.88 | 0.50 |
| 35 | 0.87 | 0.43 | 0.88 | 0.38 |
| **Average** | **0.86** | **0.51** | **0.88** | **0.58** |

**Table 3.** User-specific AUC scores in the flight game (tree depth 9).

| User | Decision Tree | | Random Forest | |
|---|---|---|---|---|
| | Train | Test | Train | Test |
| 1 | 0.90 | 0.49 | 0.85 | 0.63 |
| 2 | 0.88 | 0.84 | 0.85 | 1.00 |
| 3 | 0.86 | 0.50 | 0.84 | 0.47 |
| 4 | 0.87 | 0.57 | 0.85 | 0.70 |
| 5 | 0.88 | 0.77 | 0.85 | 0.76 |
| 6 | 0.88 | 0.60 | 0.86 | 0.71 |
| 7 | 0.89 | 0.54 | 0.86 | 0.77 |
| 8 | 0.89 | 0.74 | 0.86 | 0.87 |
| 9 | 0.88 | 0.04 | 0.84 | 0.01 |
| 10 | 0.88 | 0.64 | 0.86 | 0.77 |
| 11 | 0.90 | 0.78 | 0.85 | 0.83 |
| 12 | 0.89 | 0.66 | 0.86 | 0.57 |
| **13** | **0.88** | **0.98** | **0.85** | **1.00** |
| 14 | 0.90 | 0.52 | 0.87 | 0.53 |
| 15 | 0.89 | 0.54 | 0.87 | 0.62 |
| 16 | 0.89 | 0.59 | 0.86 | 0.60 |
| 17 | 0.88 | 0.47 | 0.86 | 0.89 |
| 18 | 0.88 | 0.64 | 0.85 | 0.12 |
| 19 | 0.90 | 0.64 | 0.87 | 0.72 |
| 20 | 0.88 | 0.67 | 0.85 | 0.99 |
| **Average** | **0.88** | **0.62** | **0.86** | **0.72** |

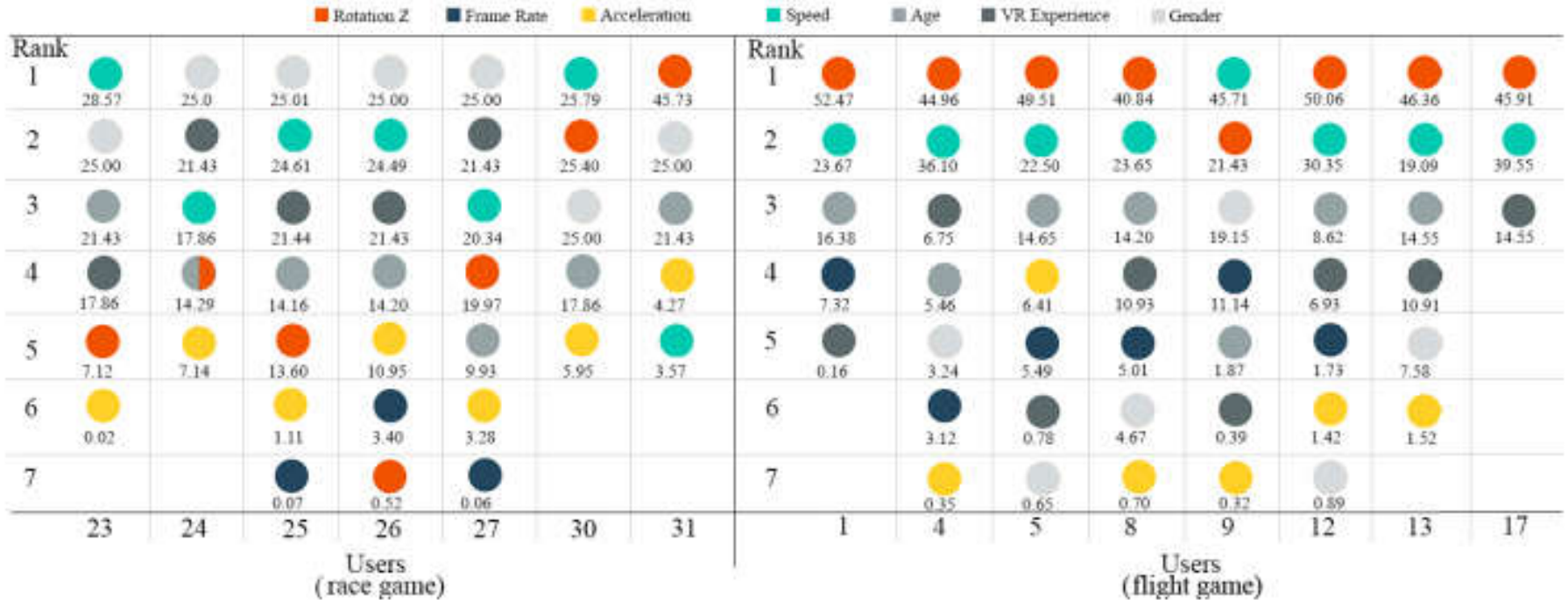

**Fig. 4.** PCS feature ranking for the race and flight games. In flight game, the rotation Z feature was ranked as the most important in 6 out of 7 participants. Moreover, it is important to highlight that the feature ranking varies substantially according to each game.

## 5 Discussion

Predicting the cause of CS is not trivial. Every user has a specific susceptibility to discomfort. Furthermore, several attributes are related to the hardware and ergonomic aspects of the devices. We are still far from tracing very precise causes for all specific cases. However, so far, factors such as rotation, speed, gender, and previous VR experience, appeared as dominant factors that may trigger CS.

Our approach works with seven factors attributed to CS. Previous works in the literature already proposed strategies for four of these attributes (acceleration, speed, frame rate, and camera rotation on the z-axis) [4, 6, 33]. The remaining causes (gender, VR experience, and age) are causes associated to the user profile and are still not associated to a clear strategy.

In addition, we observed different patterns of causes for users in the race game when compared to the flight game. In the flight case, rotation accompanied by speed were the most frequently estimated causes for the discomfort in 7 out of 8 users. Strategies to minimize CS could be applied to circumvent the influence of speed and rotation.

The race game is associated to acceleration shifts controlled by the user, whereas the flight game acceleration is nearly non existent and is not controlled by the user. This difference was very clear in the ranking results where it is possible to confirm that the CS triggered by acceleration was more prevalent in the race case rather than the flight case. In other words, the PCS average for the acceleration attributes over all users were: Race Game = 4.67, Flight Game = 1.34.

Another feature that influenced the discomfort in both scenarios was the former VR experience. Our proposed PCS outputted greater scores for the race game in contrast to the flight game, 14.79 and 6.42, respectively. In fact, the race game controllers are more complex than the flight game controllers. Intro-ducing rapid movements and related variables that are controlled by the user can potentially lead to higher incidence of cybersickness. Virtual reality games that rely on low complex controllers are a good fit for non-experienced users.

Conclusively, causes were ranked differently over the games and user profiles. CS can be triggered by different factors and their combinations, where eventually a single variable influences more than the remaining. For this reason, different combinations of strategies can be applied specifically to each user.

In addition, several works [27] propose strategies based on causes. However, none of them focuses on assisting game designers by identifying the cause in order to properly suggest an appropriate strategy. Furthermore, designers can benefit from our results in order to choose the best-fit strategy to reduce CS during gameplays according to user profile or game.

## 6 Conclusion

In this work, we propose an approach to identify potential causes for CS in different VR games using HMDs. To the best of our knowledge, this is the first work that performs this analysis in real-time, during the gameplay experience.

We considered two different scenarios for the experimental analysis and proposed two symbolic machine learning algorithms. Next, we performed a feature ranking to identify the most relevant CS causes that are specific to each user.

We observed that users with the same profile are more likely to manifest the same causes that lead to CS. Furthermore, we can also assume that the introduction of more movements and/or variables controlled by the user in the game probably lead to higher incidence of CS. VR games with low complexity controllers can be a good fit to non-experienced users.

Regarding limitations, the COVID-19 pandemic affected our experiments in terms of dataset construction. For this reason, some subjective and gameplay features, such as the player posture and locomotion, were not included in the training. Besides, some features were not well represented, such as gender, age, and former VR experience.

Future work involves including features such as posture, vision impairments, locomotion and other features to our framework using new experimental protocols. We also aim to improve the balance of the dataset as some cases were also not broadly represented, such as gender (few women), age (few elders) and experience (few participants with former VR experience). Another straightforward way is to explore other machine learning models that allow similar interpretability, such as SVM and logistic regression.

As a final remark, the raw dataset of this work and created games are published in a public domain for further reproduction and comparisons [25].

## References

1. Bernardini, F., Monard, M.C., Prati, R.: Constructing ensembles of symbolic classifiers. International Journal of Hybrid Intelligent Systems **3**(3), 159–167 (2006)

2. Bles, W., Bos, J.E., De Graaf, B., Groen, E., Wertheim, A.H.: Motion sickness: only one provocative conflict? Brain research bulletin **47**(5), 481–487 (1998)
3. Bos, J.E., Bles, W., Groen, E.L.: A theory on visually induced motion sickness. Displays **29**(2), 47–57 (2008)
4. Bouyer, G., Chellali, A., Lécuyer, A.: Inducing self-motion sensations in driving simulators using force-feedback and haptic motion. In: Virtual Reality (VR), 2017 IEEE. pp. 84–90. IEEE (2017)
5. Breiman, L.: Random forests. Machine learning **45**(1), 5–32 (2001)
6. Budhiraja, P., Miller, M.R., Modi, A.K., Forsyth, D.: Rotation blurring: Use of artificial blurring to reduce cybersickness in virtual reality first person shooters. arXiv preprint arXiv:1710.02599 (2017)
7. Calvelo, M., Piñeiro, Á., Garcia-Fandino, R.: An immersive journey to the molecular structure of sars-cov-2: Virtual reality in covid-19. Computational and Structural Biotechnology Journal (2020)
8. Creighton, R.H.: Unity 3D game development by example: A Seat-of-your-pants manual for building fun, groovy little games quickly. Packt Publishing Ltd (2010)
9. Dennison, M.S., D'Zmura, M.: Cybersickness without the wobble: experimental results speak against postural instability theory. Applied ergonomics **58**, 215–223 (2017)
10. Flach, P.: Machine Learning: The Art and Science of Algorithms That Make Sense of Data. Cambridge (2012)
11. Frank, E., Hall, M.A., Witten, I.H.: Data Mining: Practical Machine Learning Tools and Techniques. Morgan Kaufmann, 4th ed. edn. (2016)
12. Garcia-Agundez, A., Reuter, C., Becker, H., Konrad, R., Caserman, P., Miede, A., Göbel, S.: Development of a classifier to determine factors causing cybersickness in virtual reality environments. Games for health journal **8**(6), 439–444 (2019)
13. Graves, A., Mohamed, A.r., Hinton, G.: Speech recognition with deep recurrent neural networks. In: 2013 IEEE international conference on acoustics, speech and signal processing. pp. 6645–6649. IEEE (2013)
14. Huang, J., Ling, C.X.: Using auc and accuracy in evaluating learning algorithms. IEEE Transactions on knowledge and Data Engineering **17**(3), 299–310 (2005)
15. Jeong, D., Yoo, S., Yun, J.: Cybersickness analysis with EEG using deep learning algorithms. In: 2019 IEEE Conference on Virtual Reality and 3D User Interfaces (VR). pp. 827–835. IEEE (2019)
16. Jin, W., Fan, J., Gromala, D., Pasquier, P.: Automatic prediction of cybersickness for virtual reality games. In: 2018 IEEE Games, Entertainment, Media Conference (GEM). pp. 1–9. IEEE (2018)
17. Kemeny, A., Chardonnet, J.R., Colombet, F.: Getting Rid of Cybersickness: In Virtual Reality, Augmented Reality, and Simulators. Springer Nature (2020)
18. Kim, H.K., Park, J., Choi, Y., Choe, M.: Virtual reality sickness questionnaire (vrsq): Motion sickness measurement index in a virtual reality environment. Applied ergonomics **69**, 66–73 (2018)
19. Kim, J., Kim, W., Oh, H., Lee, S., Lee, S.: A deep cybersickness predictor based on brain signal analysis for virtual reality contents. In: Proc. IEEE International Conference on Computer Vision. pp. 10580–10589 (2019)
20. Kolasinski, E.M.: Simulator sickness in virtual environments. Tech. rep., DTIC Document (1995)
21. Kühnapfel, U., Cakmak, H.K., Maaß, H.: Endoscopic surgery training using virtual reality and deformable tissue simulation. Computers & graphics **24**(5), 671–682 (2000)

22. Lawrence, S., Giles, C.L., Tsoi, A.C., Back, A.D.: Face recognition: A convolutional neural-network approach. IEEE transactions on neural networks **8**(1), 98–113 (1997)
23. Maree, C., Omlin, C.W.: Towards responsible ai for financial transactions. In: 2020 IEEE Symposium Series on Computational Intelligence (SSCI). pp. 16–21. IEEE (2020)
24. Pedregosa, F., Varoquaux, G., Gramfort, A., Michel, V., Thirion, B., Grisel, O., Blondel, M., Prettenhofer, P., Weiss, R., Dubourg, V., et al.: Scikit-learn: Machine learning in python. Journal of Machine Learning Research **12**, 2825–2830 (2011)
25. Porcino, T.: Cybersickness Dataset. https://github.com/tmp1986/UFFCSData, accessed: 2021-07-07
26. Porcino, T., Rodrigues, E.O., Silva, A., Clua, E., Trevisan, D.: Using the gameplay and user data to predict and identify causes of cybersickness manifestation in virtual reality games. In: 2020 IEEE 8th International Conference on Serious Games and Applications for Health (SeGAH). pp. 1–8. IEEE (2020)
27. Porcino, T., Trevisan, D., Clua, E.: Minimizing cybersickness in head-mounted display systems: causes and strategies review. In: 2020 22nd Symposium on Virtual and Augmented Reality (SVR). pp. 154–163. IEEE (2020)
28. Rebenitsch, L., Owen, C.: Review on cybersickness in applications and visual displays. Virtual Reality **20**(2), 101–125 (2016)
29. Rebenitsch, L.R.: Cybersickness prioritization and modeling. Michigan State University (2015)
30. Rodrigues, E.O., Conci, A., Liatsis, P.: Morphological classifiers. Pattern Recognition **84**, 82–96 (2018)
31. Sak, H., Senior, A.W., Beaufays, F.: Long short-term memory recurrent neural network architectures for large scale acoustic modeling (2014)
32. Statista, A.: The statistics portal. Web site: https://www.statista.com/statistics/591181/global-augmented-virtual-reality-market-size/ (2020)
33. Van Waveren, J.: The asynchronous time warp for virtual reality on consumer hardware. In: Proc. 22nd ACM Conference on Virtual Reality Software and Technology. pp. 37–46 (2016)